\documentclass[showpacs,amsmath,amssymb,11pt]{revtex4}% Physical Review D
\usepackage{graphicx}
\usepackage{amssymb}
\usepackage{epstopdf}
\usepackage{epsfig}
\allowdisplaybreaks

\begin{document}

\def\aprge{\buildrel > \over {_{\sim}}}
\def\aprle{\buildrel < \over {_{\sim}}}

\def\etal{{\it et.~al.}}
\def\ie{{\it i.e.}}
\def\eg{{\it e.g.}}

\def\bwt{\begin{widetext}}
\def\ewt{\end{widetext}}
\def\be{\begin{equation}}
\def\ee{\end{equation}}
\def\bea{\begin{eqnarray}}
\def\eea{\end{eqnarray}}
\def\bean{\begin{eqnarray*}}
\def\eean{\end{eqnarray*}}
\def\bary{\begin{array}}
\def\eary{\end{array}}
\def\bi{\bibitem}
\def\bit{\begin{itemize}}
\def\eit{\end{itemize}}

\def\lan{\langle}
\def\ran{\rangle}
\def\lra{\leftrightarrow}
\def\la{\leftarrow}
\def\ra{\rightarrow}
\def\dash{\mbox{-}}
\def\ol{\overline}

\def\ub{\ol{u}}
\def\db{\ol{d}}
\def\sb{\ol{s}}
\def\cb{\ol{c}}

\def\re{\rm Re}
\def\im{\rm Im}

\def \b{{\cal B}}
\def \ca{{\cal A}}
\def \ko{K^0}
\def \ok{\overline{K}^0}
\def \s{\sqrt{2}}
\def \st{\sqrt{3}}
\def \sx{\sqrt{6}}
%\begin{document}
%\begin{large}
\title{{\bf Estimates for $X(4350)$ Decays from the Effective Lagrangian Approach}}
\author{Yong-Liang Ma}
\address{Department of Physics, Nagoya University, Nagoya, 464-8602, Japan.\\
Theoretical Physics Center for Science Facilities, CAS, Beijing
100049, China.}
\date{\today}
%%%%%%%%%%%%%%%%%%%%%%%%%%%%%%%
\begin{abstract}
The strong and electromagnetic decays of $X(4350)$ with quantum
numbers $J^P =0^{++}$ and $2^{++}$ have been studied by using the
effective Lagrangian approach. The coupling constant between
$X(4350)$ and $D_s^{\ast}D_{s0}^{\ast}$ is determined with the help
of the compositeness condition which means that $X(4350)$ is a bound
state of $D_s^{\ast}D_{s0}^{\ast}$. Other coupling constants applied
in the calculation are determined phenomenologically. Our numerical
results show that, using the present data within the present model,
the possibility that $X(4350)$ is a $D_s^{\ast}D_{s0}^{\ast}$
molecule can not be ruled out.
\end{abstract}

\maketitle

\section{Introduction}

Recently, a hidden charm resonance named $X(4350)$ was observed by
Belle collaboration in the analysis of the $\gamma\gamma \to \phi
J/\psi$ process~\cite{Shen:2009vs}. The mass and natural width of
this resonance are measured to be
$(4350.6^{+4.6}_{-5.1}(\rm{stat})\pm 0.7(\rm{syst}))~\hbox{MeV}/{\it
c}^2$ and $(13.3^{+17.9}_{-9.1}(\rm{stat})\pm
4.1(\rm{syst}))~\hbox{MeV}$, respectively. The product of its
two-photon decay width and branching fraction to $\phi J/\psi$ is
$(6.7^{+3.2}_{-2.4}(\rm{stat}) \pm 1.1(\rm{syst}))~\hbox{eV}$ for
$J^{PC}=0^{++}$, or $(1.5^{+0.7}_{-0.6}(\rm{stat}) \pm
0.3(\rm{syst}))~\hbox{eV}$ for $J^{PC}=2^{++}$. In literature, the
structure of $X(4350)$ has been proposed to be $c\bar{c}s\bar{s}$
teraquark state with $J^{PC} = 2^{++}$~\cite{Stancu:2009ka},
$D_{s}^{\ast}D_{s0}^{\ast}$ molecular state~\cite{Zhang:2009em} and
$P$-wave charmonium state
$\chi_{c2}^{\prime\prime}$~\cite{Liu:2009fe}. And concerning the
quantum numbers of the final states $J/\psi \phi$, $X(4350)$ can
also have quantum numbers $J^{PC} = 1^{-+}$. In
Ref.~\cite{Albuquerque:2010fm}, it was shown that $X(4350)$ cannot
be a $1^{-+}$ exotic $D_{s}^{\ast}D_{s0}^{\ast}$ molecular state. In
this paper, we will accept it as a bound state of
$D_{s}^{\ast}D_{s0}^{\ast}$ to study its strong and electromagnetic
decays in the effective Lagrangian approach in case of $J^{PC} =
0^{++}$ and $2^{++}$.

Since the mass of $X(4350)$ is about $80~$MeV below the threshold of
$D_s^{\ast}D_{s0}^{\ast}$ ($m_{_{D_{s0}^{\ast}}} = 2317.8 \pm
0.6~$MeV and $m_{_{D_s^{\ast}}} = 2112.3 \pm
0.5~$MeV~\cite{Amsler:2008zz}), it is reasonable to regard $X(4350)$
as a bound state of $D_{s0}^{\ast}D_s^{\ast}$. And because the
quantum numbers of $D_{s0}^{\ast}$ and $D_{s}^{\ast}$ are $J^P =
0^+$ and $J^{P} = 1^-$ respectively, to form a bound state with
quantum numbers $J^{PC} = 0^{++}$ or $2^{++}$, the coupling between
$X(4350)$ and its constituents should be $P-$wave. To determine the
effective coupling constant between $X(4350)$ and it constituents
$D_{s}^{\ast}D_{s0}^{\ast}$, as in our previous work (for example
Ref.~\cite{Faessler:2007gv}), we resort to the compositeness
condition $Z_{_X} = 0$($Z_{_X}$ as the wave function renormalization
constant of $X(4350)$) which was early used by nuclear
physicists~\cite{Weinberg:1962hj,Salam:1962ap} and is being widely
used by particle physicists(see the references in
\cite{Faessler:2007gv}). Recently, this method has been applied to
study the properties of some ``exotic'' hadrons
~\cite{Faessler:2007gv,Faessler:2007us,Dong:2008gb,Faessler:2008vc,Ma:2008hc,Giacosa:2007bs,Branz:2007xp,Branz:2008qm,Branz:2008ha,Branz:2008cb}
and some conclusions were yielded comparing with data. For other
interactions, we write down the general effective Lagrangian and
determine the coupling constants with help of data, theoretical
calculation, $SU(4)$ relation or the vector meson dominance (VMD).

As in our previous
work~\cite{Faessler:2007gv,Faessler:2007us,Dong:2008gb,Faessler:2008vc,Ma:2008hc,Giacosa:2007bs,Branz:2007xp,Branz:2008qm,Branz:2008ha,Branz:2008cb},
we introduce a correlation function including a scale parameter
$\Lambda_{_X}$ to illustrate the distribution of the constituents in
the bound state $X(4350)$. The parameter $\Lambda_{_X}$ is varied to
find the physical region where the data can be understood. In the
physical region of $\Lambda_{_X}$, the partial widthes for strong
and electromagnetic decays are yielded.

This paper is organized as the following: In section \ref{sec:tf} we
will provide the theoretical framework used in this paper. We will
present the analytic forms for the radiative and strong decay matrix
elements and partial widths of $X(4350)$ in section \ref{sec:decay}.
And, the last section is our numerical results and discussions.

\section{Theoretical framework}

\label{sec:tf}

In this section, we will propose the theoretical framework for the
calculation of the strong and electromagnetic decays of $X(4350)$.

\subsection{The Molecular Structure of $X(4350)$}

As was mentioned above, we regard $X(4350)$ as a $D_s^\ast
D_{s0}^{\ast}$ bound state. And concerning the experimental status,
we accept the quantum numbers of $X(4350)$ as $J^P = 0^{++}$ and
$2^{++}$. For scalar case, one can write the free lagrangian of
$X(4350)$ as
\begin{eqnarray}
{\cal L}_{\rm free}^{S}& = & \frac{1}{2}\partial_\mu X
\partial_\mu X - \frac{1}{2}m_{_X}^2X^2  \; ,
\end{eqnarray}
with $m_{_X}$ as the mass of $X(4350)$. The propagator of $X(4350)$
can be easily written as
\begin{eqnarray}
G_F(x) & = & \int\frac{d^4p}{(2\pi)^4}\frac{i}{p^2 -
m_X^2-i\epsilon} \, e^{-ip\cdot x} \; ,
\end{eqnarray}
which satisfies
\begin{eqnarray}
(\partial^2 + m^2)G_F(x) & = & -i\delta^{(4)}(x)  \; .
\end{eqnarray}
While for tensor resonance we have the free Lagrangian
as~\cite{Bellucci:1994eb}
\begin{eqnarray}
{\cal L}_{\rm free}^{T} & = & -\frac{1}{2} X_{\mu\nu}
D^{\mu\nu;\lambda\sigma} X_{\lambda\sigma}  \; ,
\end{eqnarray}
where the symmetric tensor $X_{\mu\nu} = X_{\nu\mu}$ denotes the
$J^{PC} = 2^{++}$ field for $X(4350)$ and
\begin{eqnarray}
D^{\mu\nu;\lambda\sigma} & = & (\Box +
m_{_X}^2)\Big\{\frac{1}{2}(g^{\mu\lambda}g^{\nu\sigma}+g^{\nu\lambda}g^{\mu\sigma})-g^{\mu\nu}g^{\lambda\sigma}\Big\}
\nonumber\\
& & +g^{\lambda\sigma}\partial^\mu\partial^\nu +
g^{\mu\nu}\partial^\lambda\partial^\sigma -
\frac{1}{2}(g^{\nu\sigma}\partial^\mu\partial^\lambda +
g^{\nu\lambda}\partial^\mu\partial^\sigma +
g^{\mu\sigma}\partial^\nu\partial^\lambda +
g^{\mu\lambda}\partial^\nu\partial^\sigma)  \; .
\end{eqnarray}
The propagator for $X_{\mu\nu}(4350)$ is obtained as
\begin{eqnarray}
G_{\mu\nu;\lambda\sigma}(x) & = &
\int\frac{d^4p}{(2\pi)^4}\frac{i}{p^2 -
m_{_X}^2-i\epsilon}P_{\mu\nu;\lambda\sigma} e^{-ip\cdot x}  \; ,\nonumber\\
P_{\mu\nu;\lambda\sigma} & = &
\frac{1}{2}(P_{\mu\lambda}P_{\nu\sigma}+P_{\mu\sigma}P_{\nu\lambda})
- \frac{1}{3}P_{\mu\nu}P_{\lambda\sigma}  \, , \nonumber\\
P_{\mu\nu} & = & - g_{\mu\nu} + \frac{p_\mu p_\nu}{m_{_X}^2}  \; ,\nonumber\\
D^{\mu\nu;\lambda\sigma}G_{\lambda\sigma}^{\,\,\,\,\,\, \alpha\beta}
& = & -i\frac{1}{2}(g^{\mu\alpha}g^{\nu\beta} +
g^{\nu\alpha}g^{\mu\beta})\delta^{(4)}(x) \; .
\end{eqnarray}

With respect to the discussions given in first section, one can
write the effective Lagrangian describing the interaction between
$X(4350)$ and $D_s^\ast D_{s0}^{\ast}$ as
\begin{eqnarray}
{\cal L}_{\rm int}^{S} & = & \frac{i}{\sqrt{2}}g_{_S}X(x)\int dx_1
dx_2
C_{\mu\mu}(x_1,x_2)\Phi_X((x_1-x_2)^2)\delta(x-\omega_v x_1 - \omega_sx_2)  \;,\nonumber\\
%%%%%%%%%%%%%%%%%%%%%%%%%%%%%%%%%%%%%%%%%%%%%%%%%%%%%%%%%%%%%%%%%%%%%%%%%%%
{\cal L}_{\rm int}^{T} & = &
\frac{i}{\sqrt{2}}g_{_T}X^{\mu\nu}(x)\int dx_1
dx_2 \Big[ C_{\mu\nu}(x_1,x_2) + C_{\nu\mu}(x_1,x_2) - \frac{1}{4}g_{\mu\nu}C_{\alpha\alpha}(x_1,x_2)\Big]\nonumber\\
& & \;\;\;\;\;\;\;\;\;\;\;\;\;\;\;\;\;\;\;\;\;\; \times
\Phi_X((x_1-x_2)^2)\delta(x-\omega_v x_1 - \omega_sx_2)  \; ,
\label{effelcomp}
\end{eqnarray}
where ${\cal L}_{\rm int}^{S}$ is for scalar resonance case while
${\cal L}_{\rm int}^{T}$ is for tensor resonance case. $g_{_S}$ and
$g_{_T}$ are the effective coupling constants for the interaction
between $X(4350)$ and $D_s^{\ast}D_{s0}^{\ast}$ in scalar and tensor
resonance cases, respectively. $\omega_v$ and $\omega_s$ are mass
ratios which are defined as
\begin{eqnarray}
\omega_v & = & \frac{m_{_{D_s^\ast}}}{m_{_{D_s^\ast}} +
m_{_{D_{s0}^{\ast}}}} \; , \,\,\,\, \omega_s =
\frac{m_{_{D_{s0}^{\ast}}}}{m_{_{D_s^\ast}} + m_{_{D_{s0}^{\ast}}}}
\; .
\end{eqnarray}
$\Phi_X((x_1-x_2)^2)$ is a correlation function which illustrates
the distribution of the constituents in the bound state. Fourier
transform of the correlation function reads
\begin{eqnarray}
\Phi_X(y^2) & = &
\int\frac{d^4p}{(2\pi)^4}\tilde{\Phi}_X(p^2)e^{-ip\cdot y} \; .
\end{eqnarray}
To write down Lagrangian (\ref{effelcomp}), for simplicity, we have
defined the tensor $C_{\mu\nu}$ as a function of the constituents
with the explicit form
\begin{eqnarray}
C_{\mu\nu}(x_1,x_2) & = & D_{s;\,\mu}^{\ast \, +}(x_1)\partial_\nu
D_{s0}^{\ast \, -}(x_2) + D_{s;\,\nu}^{\ast \, -}(x_1)\partial_\mu
D_{s0}^{\ast \, +}(x_2) \; .
\end{eqnarray}

\begin{figure}[htbp]
\begin{center}
\includegraphics[scale=0.8]{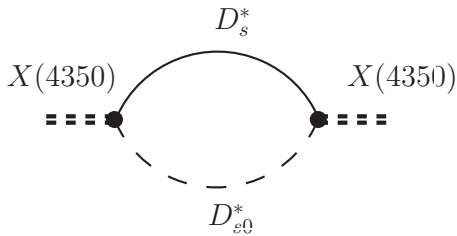}
\end{center}
\caption[The mass operator for $X(4350)$.]{%
The mass operator for $X(4350)$. } \label{fig:MassOperator}
\end{figure}

The coupling constants $g_{_S}$ and $g_{_T}$ can be determined with
help of the compositeness condition $Z_X = 0$ with $Z_X$ as the wave
function renormalization constant of $X(4350)$ which is defined as
the residual of $X(4350)$ propagator, i.e.,
\begin{eqnarray}
Z_X & = & 1 - g_X^2\frac{d}{dp^2}\Sigma_X(p^2)\Big|_{p^2=m_{_X}^2}
\; ,
\end{eqnarray}
where $g_{_X} = g_{_S}$ for scalar case while $g_{_X} = g_{_T}$ for
tensor case. For scalar resonance $X(4350)$,
$g_{_S}^2\Sigma_{_S}(p^2) = \Pi_{_S}(p^2)$ is its mass operator. But
for tensor resonance $X(4350)$, $g_{_T}^2\Sigma_{_T}(p^2)$ relates
to its mass operator via relation
\begin{eqnarray}
\Pi_{_T}^{\mu\nu;\alpha\beta}(p^2) & = &
\frac{1}{2}(g_{\mu\alpha}g_{\nu\beta} +
g_{\mu\beta}g_{\nu\alpha})g_{_T}^2\Sigma_{_T}(p^2) + \cdots \; ,
\end{eqnarray}
where $``\cdots"$ denotes terms do not contribute to the mass
renormalization of $X(4350)$. The mass operator of $X(4350)$ is
illustrated by Fig.~\ref{fig:MassOperator}.

Concerning the Feynman diagram depicted Fig.~\ref{fig:MassOperator}
one can calculate the mass operator explicitly. To get the numerical
result of the coupling constant $g_{_X}$, an explicit form of
$\tilde{\Phi}_X(p^2)$ is necessary. Throughout this paper, we take
the Gaussian form
\begin{eqnarray}
\tilde{\Phi}_X(p^2) & = & \exp(p^2/\Lambda_{_X}^2) \; ,
\label{gaussian}
\end{eqnarray}
where the size parameter $\Lambda_{_X}$ parametrizes the
distribution of the constituents inside the molecule. In the
following calculation, we will find the physical value of
$\Lambda_{_X}$ by comparing our calculation of the product of
$X(4350)$ to two-photon partial width and the branching fraction to
$J/\psi\phi$. It should be noted that choice (\ref{gaussian}) is not
unique. In principle any choice of $\tilde{\Phi}_X(p^2)$, as long as
it renders the integral convergent sufficiently fast in the
ultraviolet region, is reasonable. In this sense,
$\tilde{\Phi}_X(p^2)$ can be regarded as a regulator which makes the
ultraviolet divergent integral well defined.

\begin{figure}[htbp]
\begin{center}
\includegraphics[scale=0.8]{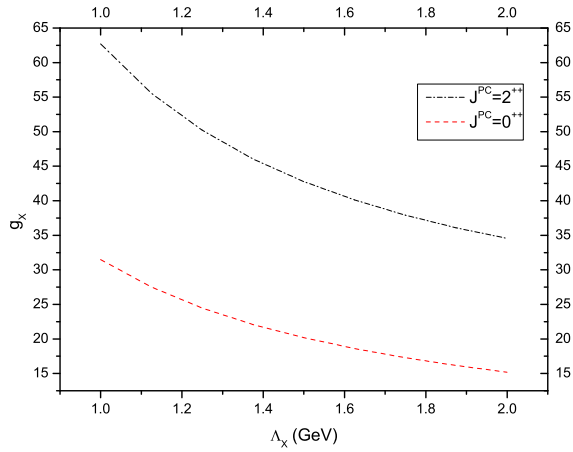}
\end{center}
\caption[$\Lambda_{_X}$ dependence of the coupling constant $g_{_X}$.]{%
$\Lambda_{_X}$ dependence of the coupling constant $g_{_X}$. }
\label{fig:gxlambda}
\end{figure}

With these discussions, we can calculate the effective coupling
constant $g_{_X}$ numerically. In the typical nonperturbative region
$\Lambda_{_X} = 1.0 \sim 2.0~$GeV, using the central value of
$X(4350)$ mass, our numerical result is found to be
%calculation. After standard
%derivation and using the central value of $X(4350)$ mass, the
%$\Lambda_{_X}$ dependence of the coupling constant $g_{_X}$ between
%$\Lambda_{_X} = 1.0 \sim 2.0~$GeV can be plotted by
%Fig.~\ref{fig:gxlambda}. That is,
\begin{eqnarray}
g_{_S} & = & 31.49 \sim 15.19 \; , \;\;\;\;\; g_{_T} = 62.70 \sim
34.54 \; .
\end{eqnarray}
%Both results decrease against $\Lambda_{_X}$.
In Fig.~\ref{fig:gxlambda} we plot the $\Lambda_{_X}$ dependence of
the coupling constants. One can see that both coupling constants
decrease against $\Lambda_{_X}$. This can be understood from the
momentum integral of the mass operator. For scalar $X(4350)$, the
loop integral is quadratically divergent so the derivative of the
mass operator which proportional to the inverse of $g_{_S}^2$
increases against $\Lambda_{_X}$ which means the coupling constant
$g_{_S}$ decreases against $\Lambda_X$. Similar argument can be
given for $g_{_T}$.

\subsection{Effective Lagrangian for Strong and Electromagnetic Decays of $X(4350)$}

The effective Lagrangian for the study of strong and electromagnetic
decays of $X(4350)$ consists of two parts: the electromagnetic part
${\cal L}_{\rm em}$ and the strong part ${\cal L}_{\rm str}$.

The electromagnetic interaction Lagrangian ${\cal L}_{\rm em}$
includes five parts: ${\cal L}_{\rm em}^{\rm NL}$ from the gauge of
the nonlocal and derivative coupling of Eq.~(\ref{effelcomp}),
${\cal L}_{\rm em}^{\rm gauge}$ from the gauge of the kinetic term
of the charged constituents $D_{s0}^{\ast}$ and $D_{s}^{\ast}$, the
electromagnetic interaction Lagrangian ${\cal L}_{\rm em}^{SV}$
including $D_{s0}^{\ast}$ and $D_s^{\ast}$, ${\cal L}_{\rm em}^{AV}$
for electromagnetic interaction including $D_{s1}$ and $D_s^{\ast}$
and ${\cal L}_{\rm em}^{AS}$ for electromagnetic interaction
including $D_{s1}$ and $D_{s0}^{\ast}$.

One can write ${\cal L}_{\rm em}^{\rm NL}$ by substituting
$C_{\mu\nu}$ with $C_{\mu\nu}^{\,\rm gauge}$ in
Eq.~(\ref{effelcomp}) with
\begin{eqnarray}
C_{\mu\nu}^{\,\rm gauge}(x_1,x_2) & = & e^{-ie I(x_1,x_2;P)}
D_{s;\,\mu}^{\ast \, +}(x_1)(\partial_\nu + i e A_\nu(x_2))
D_{s0}^{\ast \, -}(x_2) \nonumber\\
& & + e^{ie I(x_1,x_2;P)} D_{s;\,\nu}^{\ast \, -}(x_1)(\partial_\mu
- i e A_\mu(x_2)) D_{s0}^{\ast \, +}(x_2) \; ,
\end{eqnarray}
where the Wilson's line $I(x,y,P)$ is defined as
\begin{eqnarray}
I(x,y;P) & = & \int_y^x dz_\mu A^\mu(z) \; .
\end{eqnarray}
In our following calculation, the nonlocal vertex with one-photon is
necessary. The nonlocal vertex with one-photon comes from two
sources: One is from covariant derivative and another one is from
the expansion of the Wilson's line. One can easily derive the
Feynman rule for the nonlocal vertex with one-photon which comes
from the covariant derivative. But to derive the Feynman rule for
photon from Wilson's line, one may use the path-independent
prescription suggested in~\cite{Mandelstam:1962us,Terning:1991yt}.

The electromagnetic vertex ${\cal L}_{\rm em}^{\rm gauge}$ from the
gauge of the kinetic terms of the charged constituents can be easily
written as
\begin{eqnarray}
{\cal L}_{\rm em}^{\rm gauge} & = & ie A_\mu (D_{s0}^{\ast \,
-}\partial^{^{^{\hspace{-0.2cm}\leftrightarrow}}}_\mu D_{s0}^{\ast
\, +})  + ieA_\mu [ - D_{s ; \, \alpha}^{\ast \,
-}\partial^{^{^{\hspace{-0.2cm}\leftrightarrow}}}_\mu D_{s ; \,
\alpha}^{\ast \, +} + \frac{1}{2}D_{s ; \, \alpha}^{\ast \,
-}\partial^{^{^{\hspace{-0.2cm}\leftrightarrow}}}_\alpha D_{s ; \,
\mu}^{\ast \, +} + \frac{1}{2}D_{s ; \, \mu}^{\ast \,
-}\partial^{^{^{\hspace{-0.2cm}\leftrightarrow}}}_\alpha D_{s ; \,
\alpha}^{\ast \, +}] \; .
\end{eqnarray}
One can generally write the effective Lagrangian ${\cal L}_{\rm
em}^{SV}$ for electromagnetic interaction including $D_{s0}^{\ast}$
and $D_s^{\ast}$ as
\begin{eqnarray} {\cal
L}_{\rm em}^{SV} & = & eg_{_{D_{s0}^{\ast}D_s^{\ast}\gamma}} [
\tilde{V}_{\mu\nu}^{-}D_{s0}^{\ast \, +} -
\tilde{V}_{\mu\nu}^{+}D_{s0}^{\ast \, -}] F_{\mu\nu} \; .
\label{effesem}
\end{eqnarray}
where $\tilde{V}_{\mu\nu}^{\pm}$ is the gauged field strength tensor
for $D_{s}^{\ast \, \pm}$ with definition $\tilde{V}_{\mu\nu}^{\pm}
= (\partial_\mu\mp ieA_\mu)D_{s ; \, \nu}^{\ast \, \pm} -
(\partial_\nu\mp ieA_\nu)D_{s ; \, \mu}^{\ast \, \pm}$. And
similarly, the general effective Lagrangian ${\cal L}_{\rm em}^{AV}$
and ${\cal L}_{\rm em}^{AS}$ can be written as
\begin{eqnarray}
{\cal L}_{\rm em}^{AV} & = & eg_{D_{s1} D_s^\ast
\gamma}\epsilon_{\mu\nu\alpha\beta}[D_{s1;\, \mu}^- D_{s ; \,
\nu}^{\ast \, +} - D_{s1; \, \mu}^+
D_{s ;\, \nu}^{\ast \, -}]F_{\alpha\beta} \; , \nonumber\\
{\cal L}_{\rm em}^{AS} & = &  - i
eg_{D_{s1}D_{s0}^{\ast}\gamma}\epsilon_{\mu\nu\alpha\beta}
[D_{s0}^{\ast \, +}\tilde{D}_{s1 ;\,\mu\nu}^{-} - D_{s0}^{\ast \, -}
\tilde{D}_{s1; \,\mu\nu}^{+}] F_{\alpha \beta} \; .\label{effeaem}
\end{eqnarray}
Similar as the definition of $\tilde{V}_{\mu\nu}^{\pm}$, we have
defined the gauged field strength tensor for $D_{s1}^{ \pm}$ with
definition $\tilde{D}_{s1; \, \mu\nu}^{\pm} = (\partial_\mu\mp
ieA_\mu)D_{s1 ; \, \nu}^{\ast \, \pm} - (\partial_\nu\mp
ieA_\nu)D_{s1 ; \, \mu}^{\ast \, \pm}$.

The relevant coupling constants can be determined
phenomenologically. Confined by the experimental status, one cannot
fix $g_{_{D_{s0}^{\ast}D_s^{\ast}\gamma}}$ from data, so we turn to
the theoretical calculations (for example
Ref.~\cite{Faessler:2007gv} and references therein). From
literature, one can see that the minimal result of the theoretical
calculation of $D_{s0}^{\ast} \to D_s^\ast \gamma$ decay width is
$0.2~$KeV. From this decay width, we get
$g_{_{D_{s0}^{\ast}D_s^{\ast}\gamma}} \geq 3.02 \times
10^{-2}~$GeV$^{-1}$.

The coupling constants $g_{D_{s1} D_s^\ast \gamma}$ and
$g_{D_{s1}D_{s0}^{\ast}\gamma}$ can be determined by using the HQET
and branching ratio for relevant processes. First, consider the
decay of $D_{s1} \to D_s \gamma$, the effective Lagrangian can be
written as
\begin{eqnarray}
{\cal L}_{\rm em}^{D_{s1} D_s \gamma} & = & i e g_{D_{s1}D_s\gamma}
[
 D_s^+ D_{s1; \; \mu\nu}^- - D_s^- D_{s1; \;
\mu\nu}^+]F_{\mu\nu} \; ,
\end{eqnarray}
where $D_{s1; \; \mu\nu} = \partial_\mu D_{s1 ;\,\nu} -
\partial_\nu D_{s1 ;\, \mu}$ and $F_{\mu\nu} = \partial_\mu A_\nu - \partial_\nu
A_\mu$. From this Lagrangian, one can express the decay width as
\begin{eqnarray}
\Gamma(D_{s1} \to D_s \gamma) & = &
\frac{\alpha_{em}g_{D_{s1}D_s\gamma}^2}{6
m_{D_{s1}}^3}(m_{D_{s1}}^2-m_{D_s}^2)^3 \; .
\end{eqnarray}
The numerical result of the decay width has been evaluated by
several groups. From the references given in
Ref.~\cite{Faessler:2007us}, we see all the results are larger than
$0.6~$KeV. So that we have $g_{D_{s1}D_s\gamma} \geq 2.67 \times
10^{-2}~$GeV$^{-1}$. The coupling constant $g_{D_{s1} D_s^\ast
\gamma}$ relates to $g_{D_{s1}D_s\gamma}$ via HQET as
\begin{eqnarray}
\frac{g_{D_{s1}D_s\gamma}}{g_{D_{s1} D_s^\ast \gamma}} & = &
\frac{1}{m_{D_{s1}}}\frac{\sqrt{m_{D_s}}}{\sqrt{m_{D_s^\ast}}} =
3.92 \times 10^{-1}\mbox{~GeV}^{-1} \; ,
\end{eqnarray}
so that we have $g_{D_{s1} D_s^\ast \gamma} = 6.81 \times 10^{-2}$.
The coupling constant $g_{_{D_{s1} D_{s0}^{\ast}\gamma}}$ can be
determined by using the relevant branching ratio given in
PDG~\cite{Amsler:2008zz}. From (\ref{effeaem}) we have
\begin{eqnarray}
\Gamma(D_{s1} \to D_{s0}^{\ast} \gamma) & = &
\frac{2\alpha_{em}g_{D_{s1}D_{s0}^{\ast}\gamma}^2}{
3m_{D_{s1}}^3}(m_{D_{s1}}^2-m_{D_{s0}^{\ast}}^2)^3 \; .
\end{eqnarray}
Using the central value of the branching ratio we have
$\Gamma(D_{s1} \to D_{s0}^{\ast} \gamma)/\Gamma(D_{s1} \to
D_{s}\gamma) \simeq 0.21$ which leads to $g_{_{D_{s1}
D_{s0}^{\ast}\gamma}} = 3.53 \times 10^{-2}~$GeV$^{-1}$.

In addition to the Lagrangian (\ref{effelcomp}), the strong part
${\cal L}_{\rm str}$ involves $VVV$-type Lagrangian describing the
interaction of three vector mesons, the $SVV$-type Lagrangian
describing the interaction of one scalar meson with two vector
mesons, $SSV$-type Lagrangian describing the interaction of two
scalar mesons with one vector meson, $AVV$-type Lagrangian for the
interaction of axial-vector with two vector meosns and $ASV$-type
Lagrangian for axial-vector-scalar-vector meson interaction, i.e.,
\begin{eqnarray}
%{\cal L}_{\rm str} & = & {\cal L}_{\rm str}^{VVV} + {\cal L}_{\rm
%str}^{SVV} + {\cal L}_{\rm str}^{SSV} + {\cal L}_{\rm str}^{ASV}+ {\cal L}_{\rm str}^{AVV} \; ;\nonumber\\
%%%%%%%%%%%%%%%%%%%%%%%%%%%%%%%%%%%%%%%%%%%%%%%%%%%%%%%%%%%%%%%%%%%%%%%%%%%%%%%%%%%
{\cal L}_{\rm str}^{VVV} & = & ig_{_{\psi D_s^\ast D_s^\ast}}[D_{s ;\,\mu}^{\ast \, -}(D_{s ;\, \nu}^{\ast \, +}\partial^{^{^{\hspace{-0.2cm}\leftrightarrow}}}_\mu \psi_\nu) + D_{s ;\, \mu}^{\ast \, +}(\psi_\nu\partial^{^{^{\hspace{-0.2cm}\leftrightarrow}}}_\mu D_{s ;\, \nu}^{\ast \, -} ) + \psi_\mu (D_{s ;\, \nu}^{\ast \, -}\partial^{^{^{\hspace{-0.2cm}\leftrightarrow}}}_\mu D_{s ;\, \nu}^{\ast \, +})]\nonumber\\
& & + ig_{_{\phi D_s^\ast D_s^\ast}}[D_{s ;\,\mu}^{\ast \, -}(D_{s ;\, \nu}^{\ast \, +}\partial^{^{^{\hspace{-0.2cm}\leftrightarrow}}}_\mu \phi_\nu) + D_{s ;\, \mu}^{\ast \, +}(\phi_\nu\partial^{^{^{\hspace{-0.2cm}\leftrightarrow}}}_\mu D_{s ;\, \nu}^{\ast \, -} ) + \phi_\mu (D_{s ;\, \nu}^{\ast \, -}\partial^{^{^{\hspace{-0.2cm}\leftrightarrow}}}_\mu D_{s ;\, \nu}^{\ast \, +})] \; ,\\
%%%%%%%%%%%%%%%%%%%%%%%%%%%%%%%%%%%%%%%%%%%%%%%%%%%%%%%%%%%%%%%%%%%%%%%%%%%%%%%%%%%
{\cal L}_{\rm str}^{SVV} & = & g_{_{\psi D_{s0}^{\ast}D_s^{\ast}}}[D_{s0}^{\ast \, -}D_{s ; \,\mu\nu}^{\ast \, +} - D_{s0}^{\ast \, +}D_{s ;\,\mu\nu}^{\ast \, -}]\psi_{\mu\nu} + g_{_{\phi D_{s0}^{\ast}D_s^{\ast}}}[D_{s0}^{\ast \, -}D_{s ;\,\mu\nu}^{\ast \, +} - D_{s0}^{\ast \, +}D_{s ;\,\mu\nu}^{\ast \, -}]\phi_{\mu\nu} \; , \\
%%%%%%%%%%%%%%%%%%%%%%%%%%%%%%%%%%%%%%%%%%%%%%%%%%%%%%%%%%%%%%%%%%%%%%%%%%%%%%%%%%%
{\cal L}_{\rm str}^{SSV} & = & - ig_{_{\psi D_{s0}^{\ast}
D_{s0}^{\ast}}}\psi_\mu(D_{s0}^{\ast \,
-}\partial^{^{^{\hspace{-0.2cm}\leftrightarrow}}}_\mu D_{s0}^{\ast
\, +}) - ig_{_{\phi D_{s0}^{\ast}
D_{s0}^{\ast}}}\phi_\mu(D_{s0}^{\ast \,
-}\partial^{^{^{\hspace{-0.2cm}\leftrightarrow}}}_\mu D_{s0}^{\ast
\, +}) \; , \\
%%%%%%%%%%%%%%%%%%%%%%%%%%%%%%%%%%%%%%%%%%%%%%%%%%%%%%%%%%%%%%%%%%%%%%%%%%%%%%%%%%%
{\cal L}_{\rm str}^{AVV} & = & - g_{\psi D_s^\ast
D_{s1}}\epsilon_{\mu\nu\alpha\beta}[ D_{s1 ;\,\mu}^- D_{s ;\,
\nu}^{\ast \, +} - D_{s1 ;\,\mu}^+ D_{s ; \,
\nu}^{\ast \, -}]\psi_{\alpha\beta} \nonumber\\
& & - g_{\phi D_s^\ast D_{s1}}\epsilon_{\mu\nu\alpha\beta}[ D_{s1
;\,\mu}^- D_{s ;\, \nu}^{\ast \, +} - D_{s1 ;\,\mu}^+ D_{s ;\,
\nu}^{\ast \, -}]\phi_{\alpha\beta} \; , \\
%%%%%%%%%%%%%%%%%%%%%%%%%%%%%%%%%%%%%%%%%%%%%%%%%%%%%%%%%%%%%%%%%%%%%%%%%%%%%%%%%%%
{\cal L}_{\rm str}^{ASV} & = & - i g_{\psi D_{s0}^\ast D_{s1}}\epsilon_{\mu\nu\lambda\sigma}[D_{s0}^{\ast \, -}\psi_{\mu\nu}D_{s1 ;\,\lambda\sigma}^+ - D_{s0}^{\ast \, +}\psi_{\mu\nu}D_{s1 ;\,\lambda\sigma}^-]\nonumber\\
& & - i g_{\phi D_{s0}^\ast
D_{s1}}\epsilon_{\mu\nu\lambda\sigma}[D_{s0}^{\ast \,
-}\phi_{\mu\nu}D_{s1 ;\,\lambda\sigma}^+ - D_{s0}^{\ast \,
+}\phi_{\mu\nu}D_{s1 ;\,\lambda\sigma}^-] \; . \label{effelast}
\end{eqnarray}

Because of our less knowledge, we can not determine these coupling
constants from data. Here, we resort to the vector meson dominance
(VMD) model~\cite{Klingl:1996by} .
%Here we resort to the $SU(4)$ relation. Under $SU(4)$
%invariance, we have the following ratio about the coupling constants
%\begin{eqnarray}
%\frac{g_{\phi D_s^\ast D_s^\ast}}{g_{\psi D_s^\ast D_s^\ast}} & = &
%\frac{g_{\phi D_{s0}^\ast D_{s0}^{\ast}}}{g_{\psi D_{s0}^\ast
%D_{s0}^{\ast}}} = \frac{g_{\phi D_s^\ast D_{s1}}}{g_{\psi D_s^\ast
%D_{s1}}} = \frac{\sqrt{3}}{2} \nonumber\\
%%%%%%%%%%%%%%%%%%%%%%%%%%%%%%%%%%%%%%%%%%%%%%%%%%%%%%%%%%%%%%%%%%%%%%%%%%%%
%\frac{g_{\phi D_{s0}^\ast D_s^\ast}}{g_{\psi D_{s0}^\ast D_s^\ast}}
%& = & \frac{g_{\phi D_{s0}^\ast D_{s1}}}{g_{\psi D_{s0}^\ast
%D_{s1}}} = - \sqrt{3}
%\end{eqnarray}
In the VMD model, the virtual photon in the process $e^-
D_{s0}^{\ast \, +} \to e^- D_{s0}^{\ast \, +}$ is coupled to vector
mesons $\phi$ and $J/\psi$, which are then coupled to $D_{s0}^{\ast
\, +}$. For zero momentum transfer, one has relation
\begin{eqnarray}
\sum_{V = \phi, \psi} \frac{\gamma_{_V} g_{_{V
D_{s0}^{\ast}D_{s0}^{\ast}}}}{m_{_V}^2} & = & e \; ,
\label{VMDrelation}
\end{eqnarray}
where $\gamma_{_V}$ is the photon-vector-meson mixing amplitude
\begin{eqnarray}
{\cal L}_{V-\gamma-{\rm mixing}} & = & \gamma_{_V} V_\mu A_\mu \; ,
\end{eqnarray}
which can be determined from $V \to e^+ e^-$ decay width, i.e.,
\begin{eqnarray}
\Gamma_{Vee} & = & \frac{\alpha_{em} \gamma_{_V}^2}{3m_{_V}^3} \; ,
\end{eqnarray}
where we did not include electron mass since it is much smaller than
vector meson mass. For $\phi$ meson, using $\Gamma(\phi \to e^+ e^-)
= 2.97 \times 10^{-4} \times 4.26~$MeV~\cite{Amsler:2008zz} we have
$\gamma_\phi = 23472.3~$MeV$^{2}$, while $\gamma_\psi =
259965.8~$MeV$^2$ when $\Gamma(\psi \to e^+ e^-) = 5.94\% \times
93.2~$KeV~\cite{Amsler:2008zz} is applied. Concerning that the
virtual photon sees the charge of charm quark in $D_{s0}^{\ast \,
+}$ meson through $\psi D_{s0}^{\ast}D_{s0}^{\ast}$ coupling and the
charge of anti-strange quark in $D_{s0}^{\ast \, +}$ meson through
$\phi D_{s0}^{\ast}D_{s0}^{\ast}$ coupling, we have relations
\begin{eqnarray}
\frac{\gamma_\psi g_{_{\psi D_{s0}^{\ast}D_{s0}^{\ast}}}}{m_\psi^2}
& = & \frac{2}{3}e \; , \;\;\;\;\;\; \frac{\gamma_\phi g_{_{\phi
D_{s0}^{\ast}D_{s0}^{\ast}}}}{m_\phi^2} = \frac{1}{3}e \; .
\label{psiphiVMD}
\end{eqnarray}
From these relations we have $g_{_{\psi D_{s0}^{\ast}D_{s0}^{\ast}}}
= 7.45$ and $g_{_{\phi D_{s0}^{\ast}D_{s0}^{\ast}}} = 4.47$. To
determine coupling constants $g_{_{VD_{s0}^{\ast}D_s^\ast}}$, we
make extension to the VMD model used above, i.e., substituting
Eq.~(\ref{VMDrelation}) with
\begin{eqnarray}
\sum_{V = \phi, \psi} \frac{\gamma_{_V}
g_{_{VD_{i}D_{j}}}}{m_{_V}^2} & = & e g_{_{D_{i}D_{j}\gamma}} \; ,
\label{extenVMD}
\end{eqnarray}
where $D_i$ and $D_j$ denote the relevant charmed-strange mesons.
Similarly, Eqs.~(\ref{psiphiVMD}) should also be extended to
\begin{eqnarray}
\frac{\gamma_\psi g_{\psi D_iD_j}}{m_\psi^2} & = &
\frac{2}{3}eg_{D_iD_j\gamma} \; , \;\;\;\;\;\; \frac{\gamma_\phi
g_{\phi D_iD_j}}{m_\phi^2} = \frac{1}{3}e g_{D_iD_j\gamma} \; .
\label{extenpsiphiVMD}
\end{eqnarray}
From which we yield the relevant coupling constants as
\begin{eqnarray}
g_{\psi D_{i}D_{j}} & = & 7.45 \times g_{_{D_{i}D_{j}\gamma}} \; ,
\;\;\; g_{\phi D_{i}D_{j}} = 4.47 \times g_{_{D_{i}D_{j}\gamma}} \;
.
\end{eqnarray}

To fix the magnitude of coupling constant
$g_{VD_s^{\ast}D_{s}^{\ast}}$, we resort to the $SU(4)$ relation as
was used in Ref.~\cite{Lin:1999ad} from which we have relations
\begin{eqnarray}
g_{\psi D_{s}^{\ast}D_s^\ast} & = & \frac{2}{\sqrt{3}} \, g_{\phi
D_{s}^{\ast}D_s^\ast} = g_{\psi D^{\ast}D^\ast} = 7.64 \; .
\end{eqnarray}

To fix the relative signs for the relevant effective Lagrangian, one
can use HHChPT including $D_{s0}^{\ast}$ and $D_{s1}$
mesons~\cite{Casalbuoni:1996pg}. But even with this consideration,
the relative signs of ${\cal L}_{\rm em}^{AS}$ and ${\cal L}_{\rm
str}^{ASV}$ to the other terms cannot be determined. We leave this
as an ambiguity and discuss different cases in the following
calculation. In summary, our framework of the interaction Lagrangian
is
\begin{eqnarray}
{\cal L}_{\rm int} & = & {\cal L}_{\rm em}^{\rm NL} + {\cal L}_{\rm
em}^{\rm gauge} + {\cal L}_{\rm em}^{SV} + {\cal L}_{\rm em}^{AV} +
{\cal L}_{\rm str}^{VVV} + {\cal L}_{\rm str}^{SVV} + {\cal L}_{\rm
str}^{SSV} + {\cal L}_{\rm str}^{AVV} + {\cal L}_{\rm int}^{ASV}
\, ,\\
{\cal L}_{\rm int}^{ASV} & = & \pm [ \, {\cal L}_{\rm em}^{AS} +
{\cal L}_{\rm str}^{ASV} \, ] \,. \label{lagint}
\end{eqnarray}

Up to now, we have fixed all the coupling constants that are
necessary for our following calculation of the electromagnetic and
strong decays of $X(4350)$.

\section{Electromagnetic and strong decays of $X(4350)$}

\label{sec:decay}

In this section, we will present the general forms of the matrix
elements and partial widths for the electromagnetic and strong
decays of $X(4350)$ and the Feynman diagrams included in our
calculation.

\subsection{Electromagnetic decay of $X(4350)$}

\begin{figure}[htbp]
\begin{center}
\includegraphics[scale=0.55]{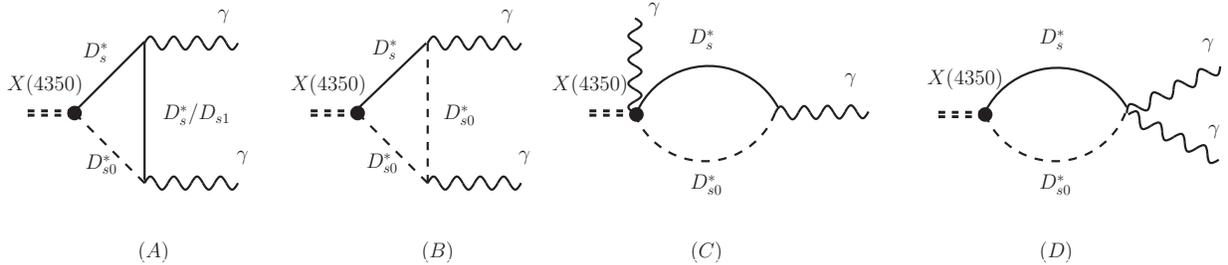}
\end{center}
\caption[Feynman diagrams for decay $X(4350) \to \gamma\gamma$(cross diagrams should be included).]{%
Feynman diagrams for decay $X(4350) \to \gamma\gamma$(cross diagrams
should be included). } \label{fig:em}
\end{figure}

The four kinds of diagrams depicted in Fig.~\ref{fig:em} and their
corresponding crossing ones should be taken into account to study
$X(4350) \to 2\gamma$ decay. Diagrams $(A)$ and $(B)$ are from the
final state interaction due to the exchange of $D_{s}^{\ast}$,
$D_{s1}$ and $D_{s0}^{\ast}$, diagram $(C)$ arises from the gauge of
the nonlocal and derivative coupling between $X(4350)$ and its
constituents $D_s^\ast D_{s0}^{\ast}$ but diagram $(D)$ is from the
Lagrangian (\ref{effesem}).

For $X(4350)$ with quantum numbers $J^{PC} = 0^{++}$, concerning the
$U(1)_{\rm em}$ gauge invariance and the transverseness of the
photon polarization vector, one can write down the matrix element
for the decay of $X \to 2\gamma$ as
\begin{eqnarray}
iM_{\rm S}^{\rm em} & = & i e^2 F_{X_s\to
2\gamma}\Big[g_{\alpha\beta} - \frac{q_{2 \, \alpha} q_{1 \,
\beta}}{q_1\cdot q_2}\Big]\epsilon_\alpha(q_1)\epsilon_\beta(q_2)
\;. \label{matrixems}
\end{eqnarray}
While for tensor meson $X(4350)$ with quantum numbers $J^{PC} =
2^{++}$, its polarization vector satisfies $\epsilon^{\mu\nu} =
\epsilon^{\nu\mu}$ and $\epsilon^\mu_\mu = 0$, so that the matrix
element for electromagnetic decay can be written
as~\cite{Anisovich:2001zp,Anisovich:2002im}
\begin{eqnarray}
iM_{\rm T}^{\rm em} & = & i e^2 \Big\{ F^{(0)}_{T \to 2\gamma} \Big[
g_{\alpha\beta} - \frac{q_{2 \, \alpha}q_{1 \, \beta}}{q_1\cdot
q_2}\Big]\frac{q_\mu q_\nu}{q^2}  \label{matrixemt}\\
& & + F^{(2)}_{T \to 2\gamma}\Big[(g_{\mu\alpha} - \frac{q_\mu q_{
\alpha}}{q^2})(g_{\nu\beta} - \frac{q_\nu q_{\beta}}{q^2}) +
(g_{\mu\beta} - \frac{q_\mu q_{\beta}}{q^2})(g_{\nu\alpha} -
\frac{q_\nu
q_{\alpha}}{q^2})\Big]\Big\}\epsilon_{\mu\nu}(p)\epsilon_\alpha(q_1)\epsilon_\beta(q_2)
\; . \nonumber
\end{eqnarray}
%with
%\begin{eqnarray}
%S_{\mu\nu,\alpha\beta}^{(0)}(p,q) & = & g_{\alpha\beta}^{\perp\perp}(\frac{q_\mu q_\nu}{q^2}-\frac{1}{3}g_{\mu\nu}^{\perp}); \nonumber\\
%S_{\mu\nu,\alpha\beta}^{(2)}(p,q) & = &
%g_{\mu\alpha}^{\perp\perp}g_{\nu\beta}^{\perp\perp} +
%g_{\mu\beta}^{\perp\perp}g_{\nu\alpha}^{\perp\perp} -
%g_{\mu\nu}^{\perp\perp}g_{\alpha\beta}^{\perp\perp} \label{tensorem}
%\end{eqnarray}
%which satisfy
%$S_{\mu\nu,\alpha\beta}^{(0)}S_{\mu^\prime\nu^\prime,\alpha\beta}^{(2)}
%= 0$. We have defined $p = q_1 + q_2$ and $q = q_1 - q_2$ and the
%metric tensors
%\begin{eqnarray}
%g_{\mu\nu}^{\perp} & = & g_{\mu\nu} - \frac{p_\mu p_\nu}{p^2};
%\;\;\;\;  g_{\mu\nu}^{\perp\perp} = g_{\mu\nu} - \frac{q_\mu
%q_\nu}{q^2} - \frac{p_\mu p_\nu}{p^2}
%\end{eqnarray}
%One can see that $g_{\mu\nu}^{\perp}$ works in the space orthogonal
%to $p$ while $g_{\mu\nu}^{\perp\perp}$ works in the space orthogonal
%to both $p$ and $q$.
where $q = q_1-q_2$. From Eqs.~(\ref{matrixems},\ref{matrixemt}) we
express the decay width for $X(4350)$ as
\begin{eqnarray}
\Gamma_{\rm S}(X \to 2\gamma) & = & \frac{2\pi}{m_X}\alpha_{em}^2F_{X_S \to 2\gamma}^2 \; , \nonumber\\
\Gamma_{\rm T}(X \to 2\gamma) & = &
\frac{\pi}{15m_X}\alpha_{em}^2(5F^{(0)2}_{X_T \to 2\gamma}
-4F^{(0)}_{X_T \to 2\gamma}F^{(2)}_{X_T \to 2\gamma} +
32F^{(2)2}_{X_T \to 2\gamma}) \; ,
\end{eqnarray}
where the subindices ``S'' and ``T'' denote the scalar and tensor
resonance $X(4350)$, respectively. To get the last equation, we have
applied the sum of the polarization vector for tensor
meson~\cite{LopezCastro:1997im}
\begin{eqnarray}
\sum_{\rm polar}
\epsilon_{\mu_1\nu_1}(p)\epsilon_{\mu_2\nu_2}^{\ast}(p) & = &
\frac{1}{2}\Big(\theta_{\mu_1\mu_2}\theta_{\nu_1\nu_2} +
\theta_{\mu_1\nu_2}\theta_{\nu_1\mu_2}\Big) -
\frac{1}{3}\theta_{\mu_1\nu_1}\theta_{\mu_2\nu_2} \; ,
\label{sumtensorvect}
\end{eqnarray}
where $\theta_{\mu\nu} = -g_{\mu\nu} + \frac{p_\mu p_\nu}{m_X^2}$.
%With these discussions and after standard calculation, we get the
%final results for the radiative decay $X(4350) \to 2\gamma$ in the
%region $\Lambda_X = 1.0 \sim 2.0~$GeV as
%\begin{eqnarray}
%\Gamma_{\rm S}(X \to 2\gamma) & = & 1.94 \sim 11.14~\mbox{MeV}; \nonumber\\
%\Gamma_{\rm T}(X \to 2\gamma) & = & 23.34 \sim 206.4~\mbox{KeV}
%\end{eqnarray}
%These results show that the electromagnetic decay width for tensor
%case is much smaller than that for scalar case.
%The dependence of
%decay width on the typical scale $\Lambda_{_X}$ is plotted in Fig.
%\begin{figure}[htbp]
%\begin{center}
%\includegraphics[scale=0.7]{widthems.eps}\includegraphics[scale=0.7]{widthemt.eps}
%\end{center}
%\caption[$\Lambda_X$ dependence of the $X \to 2\gamma$ decay width. The left panel is for scalar resonance case while the right panel is for tensor resonance case.]{%
%$\Lambda_X$ dependence of the $X \to 2\gamma$ decay width. The left
%panel is for scalar resonance case while the right panel is for
%tensor resonance case. } \label{fig:widthemlambda}
%\end{figure}

\subsection{Strong decay of $X(4350)$}

% \label{sec:str}

\begin{figure}[htbp]
\begin{center}
\includegraphics[scale=0.55]{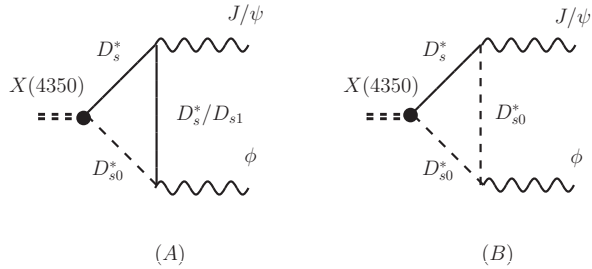}
\end{center}
\caption[Feynman diagrams for decay $X(4350) \to J/\psi \phi$(crossing diagrams should be included).]{%
Feynman diagrams for decay $X(4350) \to J/\psi \phi$(crossing
diagrams should be included). } \label{fig:strong}
\end{figure}

We should take into account the Feynman diagrams illustrated in
Fig.~\ref{fig:strong} in the study of the strong decay of $X(4350)
\to J/\psi \phi$. Furthermore, in addition to these diagrams, their
crossing ones should also be included.

Compared to the electromagnetic case, the expression for the matrix
element of strong decay is more complicated  because the constraint
from the gauge invariance is released. When $X(4350)$ is regarded as
a scalar resonance, the matrix element for the strong decay of $X
\to V_\alpha(q_1) V_\beta(q_2)$ can be written as
\begin{eqnarray}
iM_{\rm S}^{\rm str} & = & i \Big[G_{_{X_s \to V_1V_2}}
g_{\alpha\beta} + F_{_{X_s \to V_1V_2}}\frac{q_{2 \, \alpha} q_{1 \,
\beta} }{q_1 \cdot q_2}\Big]\epsilon_\alpha(q_1)\epsilon_\beta(q_2)
\; . \label{matrixstrongS}
\end{eqnarray}
One can show that, when the gauge invariance is imposed, $G_{_{X_s
\to V_1V_2}} = - F_{_{X_s \to V_1V_2}}$ so expression
(\ref{matrixstrongS}) becomes (\ref{matrixems}). Similarly, without
the constraint from the gauge invariance, in the tensor case, one
can write the matrix element for the strong decay of $X_{\mu\nu} \to
V_\alpha(q_1) V_\beta(q_2)$ as
\begin{eqnarray}
iM_{\rm T}^{\rm str} & = & i \Big[ F_{X_T \to V_1V_2}^{(1)}
g_{\alpha\beta}q_\mu q_\nu + F_{X_T \to V_1V_2}^{(2)}(g_{\mu\alpha}
g_{\nu\beta} + g_{\nu\alpha}g_{\mu\beta}) + F_{X_T \to V_1V_2}^{(3)}
(g_{\mu\alpha}q_\nu q_{1 \, \beta} + g_{\nu\alpha} q_\mu q_{1\,
\beta}) \nonumber\\
& & + F_{X_T \to V_1V_2}^{(4)} (g_{\mu\beta}q_\nu q_{2 \, \alpha} +
g_{\nu\beta} q_\mu q_{2\, \alpha}) + F_{X_T \to V_1V_2}^{(5)} q_\mu
q_\nu q_{2 \, \alpha} q_{1 \, \beta}
\Big]\epsilon_{\mu\nu}(p)\epsilon_\alpha(q_1)\epsilon_\beta(q_2) \;
.
\end{eqnarray}
One can prove that when the final vector mesons are both massless
particles and the gauge invariance is imposed the following
relations can be reduced
\begin{eqnarray}
F_{X_T \to V_1V_2}^{(3)} & = & - F_{X_T \to V_1V_2}^{(4)} = \frac{1}{2q_1\cdot q_2}F_{X_T \to V_1V_2}^{(2)} \; , \nonumber\\
F_{X_T \to V_1V_2}^{(5)} & = & -\frac{1}{q_1\cdot q_2}F_{X_T \to
V_1V_2}^{(3)} - \frac{1}{2(q_1\cdot q_2)^2}F_{X_T \to V_1V_2}^{(2)}
\; .
\end{eqnarray}
So expression (\ref{matrixemt}) for electromagnetic decay matrix
element can be yielded.

With the help of (\ref{sumtensorvect}) one can get the analytic
forms for the strong decay as
\begin{eqnarray}
\Gamma_{\rm S}(X \to J/\psi\phi) & = & \frac{1}{16\pi m_X^3}\lambda^{1/2}(m_X^2,m_\psi^2,m_\phi^2)\nonumber\\
& & \times \Big\{G_{_{X_s \to V_1V_2}}^2 \Big[2 + \omega^2\Big] -2G_{_{X_s \to V_1V_2}}F_{_{X_s \to V_1V_2}}\Big[1-\omega^2\Big] + F_{_{X_s \to V_1V_2}}^2\Big[\omega - \frac{1}{\omega}\Big]^2\Big\} \; , \nonumber\\
\Gamma_{\rm T}(X \to J/\psi\phi) & = & \frac{1}{80\pi
m_X^3}\lambda^{1/2}(m_X^2,m_\psi^2,m_\phi^2)\sum_{i\geq
j=1}^{5}\Big\{C_{ij}F_{X_T \to V_1V_2}^{(i)}F_{X_T \to
V_1V_2}^{(j)}\Big\} \; , \label{expressionwidthts}
\end{eqnarray}
where $\omega = q_1\cdot q_2/(m_\psi
m_\phi)=(m_X^2-m_\psi^2-m_\phi^2)/(2m_\psi m_\phi)$. $\lambda$ is
the K$\ddot{a}$llen function and $C_{ij}$ are functions of the
relevant masses of initial and final states which will be given in
Appendix.

\section{Numerical results and discussions}

With these discussions, the numerical calculation can be performed
via standard loop derivation. Since the magnitude of $\Lambda_{_X}$
is unknown, we vary its magnitude from $0.5~$GeV to $4.0~$GeV to
find its physical region where the data can be understood. In our
estimate, we use the cental value of the total width, i.e.,
$\Gamma_{X} = 13.3~$MeV. And, because it is difficult to determine
the relative signs between ${\cal L}_{\rm em}^{AS}$ and ${\cal
L}_{\rm str}^{ASV}$ and other terms, we will consider two cases when
we do our numerical calculation, i.e., the last two terms of
Eq.~(\ref{lagint}) give positive and negative contributions to the
total Lagrangian. Our results are summarized in
Tables.~\ref{table:scalarp} and \ref{table:scalarm}.

From the numerical results, one can see that the possibility that
$X(4350)$ is a molecular state of $D_{s0}^{\ast}D_s^{\ast}$ can not
be ruled out in our model. In the case that $X(4350)$ has quantum
numbers $J^{PC} = 0^{++}$, the physical region of $\Lambda_{_X}$ is
smaller than the tensor resonance case which means the size of
scalar $X(4350)$ is bigger than the tensor one.

We would like to point out that, because we used the minimal values
of the theoretical calculation of coupling constants
$g_{_{D_{s0}^{\ast}D_s^{\ast}\gamma}}$ and $g_{_{D_{s1}D_s\gamma}}$,
our final results about the partial widths can be regarded as lower
limit. This is an ambiguity of the present calculation. In fact, the
best way to determine these coupling constants is from data, but
because of the precision of the data, we cannot along this way. When
the magnitudes of coupling constants
$g_{_{D_{s0}^{\ast}D_s^{\ast}\gamma}}$ and $g_{_{D_{s1}D_s\gamma}}$
are improved, the theoretical results of the product of the
two-photon decay width and branch fraction to $J/\psi\phi$ should be
larger than the present conclusion. In this case, compared to the
tensor $X(4350)$, the typical region of $\Lambda_{_X}$ for scalar
resonance can be reduced to an unphysically small region so one can
first rule out the possibility of a scalar molecule.

Another ambiguity in our calculation of the product of the
two-photon decay width and branch fraction to $J/\psi\phi$ is from
the total width of $X(4350)$. Here we apply the central value, i.e.,
$\Gamma_X = 13.3~$MeV. When a larger total width is applied, the
physical region of $\Lambda_{_X}$ can be enlarged. But this does not
effect the partial widths for strong and electromagnetic decays we
predict in the corresponding region of $\Lambda_{_X}$.

Finally, we conclude that, with the present data and in the
framework our model, $X(4350)$ can be interpreted as
$D_{s0}^{\ast}D_{s}^{\ast}$ molecule.

\begin{table}

\caption{\label{table:scalarp}  Our numerical results in case of
positive sign of Eq.~(\ref{lagint}).}

\begin{tabular}{lllll}
\hline \hline \hspace*{.1cm} $J^{PC}$ \hspace*{.2cm}
& \hspace*{.1cm} $\Lambda_{_X}$(GeV) \hspace*{.1cm}& \hspace*{.3cm} Branch product(eV)\hspace*{.1cm} & \hspace*{.2cm} $\Gamma_{\rm str}$(KeV)\hspace*{.1cm} & \hspace*{.2cm} $\Gamma_{\rm em}$(KeV) \hspace*{.2cm} \\
\hline \,\, $0^{++} $ & \,\,\, $ 0.5 \sim 0.7 $ \,\,\,& \,\,\,\,\,\,\,\, $2.19 \sim 10.26$ & \,\,\,\, $100.9 \sim 174.5$\,\,\,& \,\,\,\, $0.29 \sim 0.78$\,\,\, \\
\,\, $2^{++}$ & \,\,\, $ 1.1 \sim 1.8$ \,\,\,& \,\,\,\,\,\,\,\, $1.24 \sim 2.28$ & \,\,\,\, $285.3 \sim 973.5$ \,\,\,& \,\,\,\, $0.03 \sim 0.09$ \,\,\, \\
\hline \hline
\end{tabular}
%\end{center}
\end{table}

\begin{table}

\caption{\label{table:scalarm}  Our numerical results in case of
negitive sign of Eq.~(\ref{lagint}). }

\begin{tabular}{lllll}
\hline \hline \hspace*{.1cm} $J^{PC}$ \hspace*{.2cm}
& \hspace*{.1cm} $\Lambda_{_X}$(GeV) \hspace*{.1cm}& \hspace*{.3cm} Branch product(eV)\hspace*{.1cm} & \hspace*{.2cm} $\Gamma_{\rm str}$(KeV)\hspace*{.1cm} & \hspace*{.2cm} $\Gamma_{\rm em}$(KeV) \hspace*{.2cm} \\
\hline \,\, $0^{++} $ & \,\,\, $ 0.5 \sim 0.6 $ \,\,\,& \,\,\,\,\,\,\,\, $7.21 \sim 12.74$ & \,\,\,\, $373.6 \sim 391.0$\,\,\,& \,\,\,\, $0.26 \sim 0.43$\,\,\, \\
\,\, $2^{++}$ & \,\,\, $ 1.0 \sim 1.9$ \,\,\,& \,\,\,\,\,\,\,\, $0.66 \sim 2.42$ & \,\,\,\, $166.0 \sim 915.1$ \,\,\,& \,\,\,\, $0.02 \sim 0.19$ \,\,\, \\
\hline \hline
\end{tabular}
%\end{center}
\end{table}

%%%%%%%%%%%%%%%%%%%%%%%%%%%%%%%%%%%%%%%%%%%%%%%%%%%%%%%%%%%%%%%%%%%
\appendix

\section{Explicit forms for the Functions $C_{ij}$}

In this appendix, I will present the coefficients $C_{ij}$ in
formula (\ref{expressionwidthts}).
\begin{eqnarray}
C_{11} & = & \frac{1}{24m^4m_1^2m_2^2}\Big[5\lambda^2(\lambda + 12 m_1^2m_2^2)\Big] \; , \nonumber\\
C_{12} & = & \frac{-1}{12m^4m_1^2m_2^2}\Big[\lambda\Big(5\lambda(m^2+m_1^2+m_2^2) + 24 m^2 m_1^2m_2^2\Big)\Big] \; , \nonumber\\
C_{13} & = &\frac{1}{12m^4m_1^2m_2^2}\Big[5\lambda^2\Big((m^2-m_2^2)^2-m_1^4\Big)\Big] \; , \nonumber\\
C_{14} & = &\frac{-1}{12m^4m_1^2m_2^2}\Big[5\lambda^2\Big((m^2-m_1^2)^2-m_2^4\Big)\Big] \; , \nonumber\\
C_{15} & = & \frac{1}{24m^4m_1^2m_2^2}\Big[5\lambda^3(m^2 - m_1^2 - m_2^2)\Big] \; , \nonumber\\
C_{22} & = & \frac{1}{24m^4m_1^2m_2^2}\Big[5\lambda^2+44m^2(m_1^2+m_2^2)\lambda + 528 m_1^2m_2^2 m^4 \Big] \; , \nonumber\\
C_{23} & = & \frac{-1}{12m^4m_1^2m_2^2}\Big[\lambda(5\lambda + 44 m^2m_1^2)(m^2-m_1^2 + m_2^2)\Big] \; , \nonumber\\
C_{24} & = & \frac{1}{12m^4m_1^2m_2^2}\Big[\lambda(5\lambda + 44 m^2m_2^2)(m^2-m_2^2 + m_1^2)\Big] \; , \nonumber\\
C_{25} & = & \frac{-1}{24m^4m_1^2m_2^2}\Big[5\lambda^2\Big(m^4 - (m_1^2 - m_2^2)^2\Big)\Big] \; , \nonumber\\
C_{33} & = & \frac{1}{24m^4m_1^2m_2^2}\Big[\lambda^2(5\lambda + 44 m^2m_1^2)\Big] \; , \nonumber\\
C_{34} & = & \frac{-1}{12m^4m_1^2m_2^2}\Big[5\lambda^2\Big(m^4-(m_1^2 - m_2^2)^2\Big)\Big] \; , \nonumber\\
C_{35} & = & \frac{1}{24m^4m_1^2m_2^2}\Big[5\lambda^3(m^2-m_2^2 + m_1^2)\Big] \; , \nonumber\\
C_{44} & = & \frac{1}{24m^4m_1^2m_2^2}\Big[\lambda^2(5\lambda + 44 m^2m_2^2)\Big] \; , \nonumber\\
C_{45} & = & \frac{-1}{24m^4m_1^2m_2^2}\Big[5\lambda^3(m^2-m_1^2 + m_2^2)\Big] \; , \nonumber\\
C_{55} & = & \frac{1}{96m^4m_1^2m_2^2}\Big[5\lambda^4\Big] \; ,
\end{eqnarray}
where $\lambda = \lambda(m^2,m_1^2,m_2^2)$ is the K$\ddot{a}$llen
function and $m=m_{_X}, m_1=m_{\psi}, m_2=m_{\phi}$.

%%%%%%%%%%%%%%%%%%%%%%%%%%%%%%%%%%%%%%%%%%%%%%%%%%%%%%%%%%%%%%%%%%%%%%%
\acknowledgments

\label{ACK}

I would like to thank Prof. Yu-Bing Dong from IHEP in Beijing and
Prof. M.Harada in Nagoya University for their valuable discussions
and comments. This work is supported in part by the National Science
Foundation of China (NNSFC) under grant No. 10905060 and
Grant-in-Aid for Scientific Research on Innovative Areas (No. 2104)
``Quest on New Hadrons with Variety of Flavors'' from MEXT.

%%%%%%%%%%%%%%%%%%%%%%%%%%%%%%%%%%%%%%%%%%%%%%%%%%%%%%%%%%%%%%%%%%%%%%%%%%%%%%%%%%

\end{document}